# Comment on "Unified treatment for two-center one-electron molecular integrals over Slater-type orbitals with integer and noninteger principal quantum numbers"

I.I. Guseinov

*Department of Physics, Faculty of Arts and Sciences, Onsekiz Mart University, Çanakkale, Turkey*

**Abstract**

In a recent paper Özdoğan (Z. Naturforsch, 59a(2004)743) published formulas for evaluating the two-center overlap and nuclear attraction integrals over integer and noninteger $n$ Slater type orbitals. The purpose of this article is to point out that the same formulas have previously been established by Guseinov et al. (J.Mol.Model.,8(2002)272) by using the same method. As we demonstrated in our Comment (Int.J.Quantum Chem., 91(2003)62), the expansion formula for the product of two normalized associated Legendre functions in elliptical coordinates and the expansion coefficients $a_{us}^{kk'}$ presented by Özdoğan are obtained from the use of given in our papers general formulas (I.I.Guseinov, J.Phys.,B,3(1970)1399; Phys.Rev.,A,32(1985)1864; J.Mol.Struct (Theochem),336(1995)17) by changing the summation indices. It should be noted that the published by Özdoğan results are on the use of formulas for the evaluation of multicenter integrals presented in his other papers, which are also obtained from the literature (see, e.g., C.C.J. Roothaan, J.Chem.Phys., 19 (1951) 1445; A.Lofthus, Mol. Phys., 5 (1962) 105) and our articles by changing the summation indices (see Comments: I.I.Guseinov, Commun.Theor.Phys., 38 (2002) 256; Int.J.Quantum Chem., 91 (2003) 62; J.Math.Chem., 36 (2004) 123; J.Chin.Chem.Soc., 51 (2004) 877; J. Chin. Chem. Soc., 51 (2004) 1077; I.I.Guseinov, B.A.Mamedov, Can.J.Phys., 82(2004) 205, p.206).

**Keywords:** Spater type orbitals, Two-center overlap integrals, Two-center nuclear attraction integrals, Noninteger principal quantum numbers, Molecular integrals

## I. Introduction

Özdoğan in Ref.[1] published the formulas for the calculation of two-center overlap integrals and nuclear attraction integrals of types $\langle a|(1/r_a)|b\rangle$ and $\langle a|(1/r_b)|b\rangle$. It is well known that these types of two-center nuclear attraction integrals can be expressed by two-center overlap integrals [2]. In this Comment we demonstrate that the results published by Özdoğan in two-center nuclear attraction integrals are derived from the formulas for overlap integrals. The presented in Ref[1] formulas for the two-center overlap integrals also are not original and can easily be obtained from the formulas published in our papers [3-6] by changing the summation indices. As we have shown in Comments [7-12], all of the works published by Özdoğan et al. [13-19] on the calculation of multicenter integrals are also obtained from the formulas of our articles.

## 2. Theory

It is well known that the two-center nuclear attraction integrals over integer and noninteger $n$ STOs in the lined-up coordinate systems are defined by [2]:

$$U_{nl\lambda,n'l'\lambda}(\zeta,\zeta';R) = \int \chi^*_{nl\lambda}(\zeta,\vec{r}_a)\frac{1}{r_b}\chi_{n'l'\lambda}(\zeta',\vec{r}_a)dV \tag{1}$$

$$U^{(A)}_{nl\lambda,n'l'\lambda}(\zeta,\zeta';R) = \int \chi^*_{nl\lambda}(\zeta,\vec{r}_a)\frac{1}{r_a}\chi_{n'l'\lambda}(\zeta',\vec{r}_b)dV \tag{2}$$

$$U^{(B)}_{nl\lambda,n'l'\lambda}(\zeta,\zeta';R) = \int \chi^*_{nl\lambda}(\zeta,\vec{r}_a)\frac{1}{r_b}\chi_{n'l'\lambda}(\zeta',\vec{r}_b)dV \quad , \tag{3}$$

where $\lambda = |m| = |m'|$, $R = |\vec{R}_{ab}|$, $\vec{R}_{ab} = \vec{r}_a - \vec{r}_b$ and

$$\chi_{nl\lambda}(\zeta,\vec{r}) = R_n(\zeta,r)S_{l\lambda}(\theta,\varphi), \tag{4}$$

$$R_n(\zeta,r) = (2\zeta)^{n+1/2}[\Gamma(2n+1)]^{-1/2} r^{n-1}e^{-\zeta r}. \tag{5}$$

Here, $\Gamma(2n+1)$ is the gamma function.

With the help of relation

$$\frac{1}{r}R_n(\zeta,r) = 2\zeta\left[\frac{\Gamma(2n-1)}{\Gamma(2n+1)}\right]^{1/2} R_{n-1}(\zeta,r), \tag{6}$$

it is easy to express the nuclear attraction integrals (2) and (3) through the two-center overlap integrals:

$$U^{(A)}_{nl\lambda,n'l'\lambda}(\zeta,\zeta';R) = 2\zeta\left[\frac{\Gamma(2n-1)}{\Gamma(2n+1)}\right]^{1/2} S_{n-1l\lambda,n'l'\lambda}(\zeta,\zeta';R) \tag{7}$$

$$U^{(B)}_{nl\lambda,n'l'\lambda}(\zeta,\zeta';R) = 2\zeta'\left[\frac{\Gamma(2n'-1)}{\Gamma(2n'+1)}\right]^{1/2} S_{nl\lambda,n'-1l'\lambda}(\zeta,\zeta';R), \tag{8}$$

where the two-center overlap integrals are defined by

$$S_{nl\lambda,n'l'\lambda}(\zeta,\zeta';R) = \int \chi^*_{nl\lambda}(\zeta,\vec{r}_a)\chi_{n'l'\lambda}(\zeta',\vec{r}_b)dV. \tag{9}$$

Taking into account Eq.(9) we can express the $U^{(B)}$-integral through the $U^{(A)}$-integral:

$$U^{(B)}_{nl\lambda,n'l'\lambda}(\zeta,\zeta';R) = \frac{\zeta'}{\zeta}\left[\frac{\Gamma(2n+3)\Gamma(2n'-1)}{\Gamma(2n+1)\Gamma(2n'+1)}\right]^{1/2} U^{(A)}_{n+1l\lambda,n'-1l'\lambda}(\zeta,\zeta';R). \tag{10}$$

Thus, we have only two kinds of independent two-center nuclear attraction integrals, namely, $U$ and one of $U^{(A)}$ and $U^{(B)}$ - integrals which are reduced to the two-center overlap integrals. In Ref.[1], the *U* -type nuclear attraction integrals is not discussed and, therefore, it is necessary to point out that the published by Özdoğan paper sheds no new light on the subject and that it is altogether misleading.

Now we can move on to the discussion of formulas occurring in Sections 3 and 4 of Ref.[1]. Özdoğan claims that, he has presented recently in Refs.[13, 15] (see Refs.[12, 13] in [1]) the following formula for the product of two normalized associated Legendre functions:

$$T^{l\lambda,l'\lambda}(\mu,\nu) = P_{l\lambda}\left(\frac{1+\mu\nu}{\mu+\nu}\right)P_{l'\lambda}\left(\frac{1-\mu\nu}{\mu-\nu}\right)$$
$$= \sum_{k,k'}\sum_{u,s} a_{us}^{kk'}(l\lambda,l'\lambda)\frac{(\mu\nu)^s}{(\mu+\nu)^{l-2(k+k'+\lambda-u)}(\mu-\nu)^{l'}} \quad (11)$$

We have proved in Comment [8] that the Eq.[11] is obtained from the expansion relationships [3-5]

$$T^{l\lambda,l'\lambda}(\mu,\nu) = \sum_{\alpha=-\lambda}^{l}\sum_{\beta=\lambda}^{l'}\sum_{q=0}^{\alpha+\beta} g_{\alpha\beta}^q(l\lambda,l'\lambda)\frac{(\mu\nu)^q}{(\mu+\nu)^\alpha(\mu-\nu)^\beta} \quad (12)$$

by changing the summation indices. We note that the generalized binomial coefficients $F_m(N,N')$ that occur in Eqs.(14)-(17) of Ref.[1] were introduced in our articles [3, 5].

Using above mentioned Eqs.(7) and (8) it is easy to show that the nuclear attraction integrals $I_{nl\lambda,n'l'\lambda}^{(1,0)}(\zeta,\zeta';R)$ and $I_{nl\lambda,n'l'\lambda}^{(0,1)}(\zeta,\zeta';R)$ occurring in Eq.(18) of Ref.[1] are reduced to the overlap integrals $I_{nl\lambda,n'l'\lambda}^{(0,0)}(\zeta,\zeta';R)$, therefore, Eq.(18) is not the unified formula for two-center one-electron molecular integrals. In Comments [8, 9] we demonstrated that the Eq.(18) of Ref.[1] is also obtained from our formulas by changing the summation indices. It should be noted that Eqs.(19)-(22) of Ref.[1] can also be found in our papers [3-6].

Thus, the formulas presented in Ref.[1] by Özdoğan for expansion of the product of two normalized associated Legendre functions in elliptical coordinates and two-center overlap integrals are not original and they can easily be derived from the analytical relations given in our papers by changing the summation indices (see Refs. [3-6, 8,9]). It should be noted that the symbolic results of two-center overlap integrals between different combinations of quantum numbers given in Tables 1 and 2 occurring in published by Gümüş and Özdoğan paper [19] can also be obtained from the use of established in above mentioned our papers [3-5] general formulas or presented in the literature relations for overlap integrals in terms of the products of molecular auxiliary functions $A_n$ (p) and $B_n$ (pt) (see, e.g.,Ref. [20]).